\documentclass[aps,prl,twocolumn,groupedaddress,longbibliography]{revtex4-1}

\usepackage{amsmath}
\usepackage{amssymb}
\usepackage{graphicx}

\usepackage{esvect}
\MakeRobust{\vv}

\newcommand{\ts}{\textsuperscript}

\begin{document}

\title{Electrical and Thermal Generation of Spin Currents by Magnetic Graphene}

\author{Talieh S. Ghiasi$^1$,}
\email[]{t.s.ghiasi@rug.nl}

\author{Alexey A. Kaverzin$^1$, Avalon H. Dismukes$^2$, Dennis K. de Wal$^1$,\\ Xavier Roy$^2$ and Bart J. van Wees$^1$}

\affiliation{$^1$Zernike Institute for Advanced Materials, University of Groningen, Groningen, 9747 AG, The Netherlands}
\affiliation{$^2$Department of Chemistry, Columbia University, New York, NY, USA}

\begin{abstract}
The demand for compact, high-speed and energy-saving circuitry urges higher efficiency of spintronic devices that can offer a viable alternative for the current electronics. The route towards this goal suggests implementing two-dimensional (2D) materials that provide large spin polarization of charge current together with long-distance transfer of the spin information. Here, for the first time, we experimentally demonstrate a large spin polarization of the graphene conductivity ($\approx 14\%$) arising from a strong induced exchange interaction in proximity to a 2D layered antiferromagnet. 
The strong coupling of charge and spin currents in graphene with high efficiency of spin current generation, comparable to that of metallic ferromagnets, together with the observation of spin-dependent Seebeck and anomalous Hall effects, all consistently confirm the magnetic nature of graphene. The high sensitivity of spin transport in graphene to the magnetization of the outermost layer of the adjacent interlayer antiferromagnet, also provides a tool to read out a single magnetic sub-lattice. The first time observations of the electrical/thermal generation of spin currents by magnetic graphene suggest it as the ultimate building block for ultra-thin magnetic memory and sensory devices, combining gate tunable spin-dependent conductivity, long-distance spin transport and spin-orbit coupling all in a single 2D material.
\end{abstract}

\maketitle



Memory technology has been revolutionized by the discovery of the giant magnetoresistance (GMR) \cite{baibich1988giant,binasch1989enhanced} and spin transfer torque (STT) effects \cite{slonczewski1996current, myers1999current}, employed in hard disk drives (HDD) and magnetic random access memories (MRAM). These effects arise from the efficient coupling of charge and spin currents in ferromagnetic (FM) materials. This is a crucial aspect for the development of magneto-electronic devices, such as spin-valves that consist of two layers of FM materials separated by a non-magnetic layer, where altering the relative magnetization orientation of the FM layers results in a significant change in resistance \cite{vzutic2004spintronics}. The design of spin-valves, however, requires further innovation to reach the 2D limit in the miniaturization process of the prospective ultra-compact systems.  

In this regard, the architecture of atomically thin van der Waals (vdW) heterostructures not only qualifies for the low-dimensionality but also provides exceptional functionalities by integrating the unique properties of the individual layers \cite{geim2013van}. 
The most established building block for the 2D vdW hybrids is graphene as it provides a high charge carrier mobility with gate-tunability of the carrier density, in which the absence of hyperfine interactions and small intrinsic spin-orbit coupling (SOC) allow for long-distance spin transport \cite{abergel2010properties, Han2014}. If graphene is considered in combination with the 2D magnetic materials \cite{gong2017discovery, gong2019two} in particular, it would bring the technology of ultra-thin spin-logic devices to an ultimate stage where the magnetic behavior of an individual atomic layer directly modulates information transfer by the spins in the neighboring graphene layer. 

\begin{figure*} [!htb] 
\centering\includegraphics[width=0.7\textwidth]{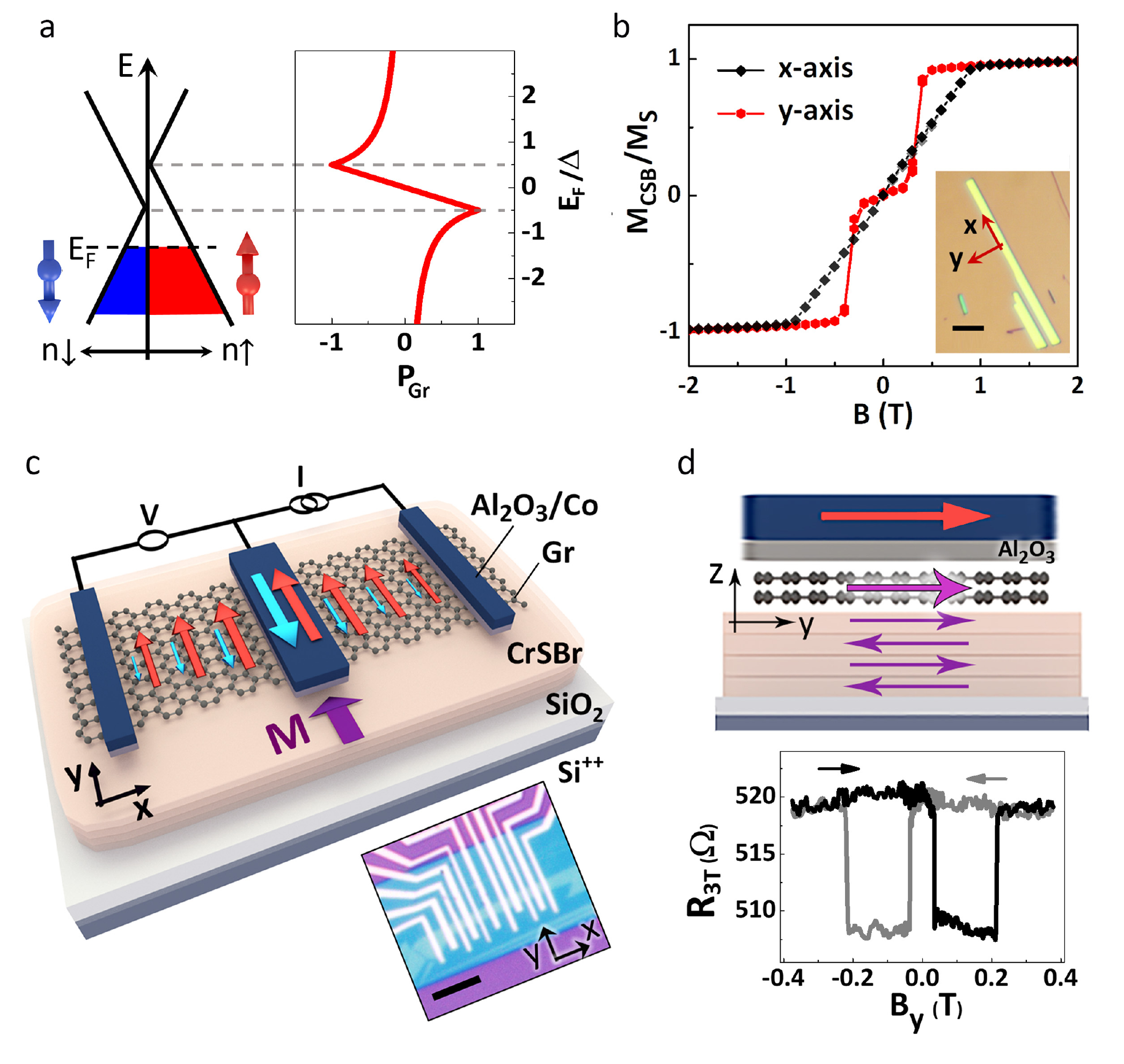}
\caption{\textbf{Induced magnetism in graphene by the proximity of chromium sulfide bromide (CrSBr).} (a) Left: Band structure (spin-dependent density of charge carriers ($n_\uparrow$ and $n_\downarrow$) versus energy) of graphene in proximity of CrSBr, with the spin-splitting $\Delta$ caused by the exchange interaction. $\Delta$ is considered to be constant for the electron and hole states around the neutrality point. Right: Dependence of spin polarization of graphene conductivity ($P_\mathrm{Gr}$) on the position of Fermi energy ($E_\mathrm{F}$), under assumptions of the Drude model (see SI, sec. 11). (b) Magnetization of a bulk CrSBr single crystal ($M_\mathrm{CSB}$) versus external magnetic field ($B$), measured using SQUID magnetometry at $T$= 30 K along the x- and y-directions of the crystal. $M_\mathrm{CSB}$ is normalized to the saturation magnetization ($M_\mathrm{s}$) value taken at $B=\,3\,\mathrm{T}$. Inset: An optical micrograph of typical CrSBr flakes, exfoliated on $\mathrm{SiO_2}$ substrate.  The x and y directions are the CrSBr crystallographic a- and b-axes, respectively. Scale bars in OM images: 5 $\mu$m. (c) Schematics of a three-terminal (3T) spin-valve measurement geometry, showing Gr/bulk CrSBr heterostructure and FM electrodes ($\mathrm{Al_2O_3}\,/\mathrm{Co}$). The red and blue arrows represent the two possible magnetization states of the Co electrode and the corresponding spins injected into the graphene channel, considering $P_\mathrm{Gr}>0$. The optical micrograph shows device D1, fabricated with a vdW stack of bilayer graphene and CrSBr ($\sim 20$ nm) and $\mathrm{Al_2O_3}$ (0.8 nm)/Co (30 nm) electrodes. (d) The sketch is the side view of the spin-valve device. The purple arrows represent the magnetization direction of the CrSBr layers, aligned with the easy y-axis. Independent switching of the magnetization directions of CrSBr and Co under an external magnetic field along y axis ($B_\mathrm{y}$) leads to the unconventional spin-valve measurements in the 3T geometry with the distinct levels of the 3T resistance $R_\mathrm{3T}=V/I$ (also see SI, section 3). The 3T measurement is performed at $T=4.5\,$K, with $I=5\,\mu$A.}
    \label{fig:one}
\end{figure*}

The 2D nature of graphene allows for efficient modulation of its band structure and modification of its charge and spin transport properties via the proximity to other materials, while its high quality of charge transport is preserved. By proximity effects, considerable spin-orbit coupling (SOC) \cite{Gmitra2015, garcia2018spin} and exchange interaction \cite{wei2016strong,wu2017magnetic, tang2020magnetic, wang2015proximity, tang2018approaching, mendes2015spin,leutenantsmeyer2016proximity, singh2017strong, karpiak2019magnetic, haugen2008spin, yang2013proximity, zollner2016theory, asshoff2017magnetoresistance,   behera2019proximity} can be induced in graphene which are essential for generation and manipulation of the spins. The induced spin-orbit and exchange interactions together with the exceptionally long spin relaxation length \cite{tombros2007electronic} make graphene an excellent material in the field of spintronics in which  a broad range of spin-dependent phenomena can be addressed \cite{fabian2007semiconductor,Han2014}. Magnetized graphene is also a perfect channel for exploration of quantum phenomena in magneto-electronic devices \cite{zhang2005experimental, tse2011quantum}. Moreover, it is an ideal component for 2D magnetic tunnel junctions (MTJ) where the generated spin currents by the magnetic graphene can induce spin-transfer torque (STT) that is applicable in 2D STT-MRAMs \cite{zhou2010magnetotransport,chappert2010emergence}.

The experimental realization of the proximity-induced exchange interaction in graphene has been reported, measuring Shubnikov de-Haas oscillations and Zeeman spin Hall effect \cite{wei2016strong, wu2017magnetic, tang2020magnetic}, anomalous Hall effect (AHE) \cite{wang2015proximity,tang2018approaching} and Hanle precession of injected spins by the induced exchange field ($\vv{B}_\mathrm{exch}$) \cite{leutenantsmeyer2016proximity, singh2017strong, karpiak2019magnetic}. Among them, the latter is the most unambiguous and reliable approach when spin-sensitive Co electrodes are used to directly detect the modulation of the spin signal by $\vv{B}_\mathrm{exch}$. However, these spin transport measurements in graphene so far have shown rather weak exchange interaction, only leading to an additional precession of the spins around the $\vv{B}_\mathrm{exch}$.  

In this work, for the first time, we detect strong spin-dependent conductivity in a proximity-induced magnetic graphene, as a result of at least an order of magnitude larger $B_\mathrm{exch}$. The unambiguous direct measurement of the spin polarization of conductance in the graphene is evidenced by active electrical and thermal generation of spins by the magnetic graphene, irrespective of spin injection by Co electrodes. These results, accompanied with the AHE measurements, not only give unique insight into the magnetic nature of graphene and its spin dynamics, but also assure its significant role in 2D spintronic and spin caloritronic circuitry, where electrical and thermal spin injection/detection is all possible only by graphene itself.

We show that the adsorbate-free interface of the vdW heterostructure of (bilayer) graphene with the 2D inter-layer antiferromagnetic (AFM) chromium sulfide bromide (CrSBr) provides the graphene channel with a strong exchange interaction, resulting in a considerable spin-splitting ($\Delta$) of the graphene band structure (Figure 1a). The resulting substantial difference in the density of charge carriers ($n$) with spin parallel ($\uparrow$) and anti-parallel ($\downarrow$) to the $\vv{B}_\mathrm{exch}$ leads to the spin-dependent conductivity. The spin polarization of conductance in graphene is expected to be efficiently tunable by shifting the position of the Fermi energy with a gate electric field (as proposed in panel a) which is the basis for an all-electric spin field effect transistors in spin-logic circuitries \cite{behin2010proposal}. The use of bilayer graphene is particularly encouraged, as it can allow for gate-tunability of the exchange splitting \cite{michetti2010electric, michetti2011spintronics,zollner2018electrically}.

The appropriate choice of the 2D magnet in such vdW heterostructures is crucial
, as most of the explored 2D magnetic materials \cite{gibertini2019magnetic} suffer from extreme instability in air and a low temperature of magnetic transitions. Here we tackle this major obstacle by utilizing the recently explored CrSBr 2D crystal that is an air-stable vdW semiconductor (bandgap $\sim 1.5\,$eV) with an interlayer AFM ordering up to a relatively high N\'eel temperature of $T_\mathrm{N}$ $\approx 132\,\mathrm{K}$ \cite{goser1990magnetic, qi2018electrically,telford2020layered}. Furthermore, the antiferromagnetism potentially promises ultra-fast operations and robustness against external magnetic fields \cite{jungwirth2016antiferromagnetic} and also is expected to be tunable by a gate electric field \cite{qi2018electrically, jiang2018electric}. Mechanical cleavage of the CrSBr crystal results in flakes with a specific rectangular geometry that correlates with its in-plane magnetic anisotropy axes 
(inset of Figure 1b). 
The magnetization behavior of CrSBr ($\vv{M}_\mathrm{CSB}$) measured versus external magnetic field ($B$), using SQUID magnetometry (Figure 1b), displays a sharp modulation of $\vv{M}_\mathrm{CSB}$ when $B$ is applied along the y-axis. 
This corresponds to the AFM behavior with spin-flip transition (at $B_\mathrm{M,y}\sim 0.2\,\mathrm{T}$) and defines the y-direction as the in-plane magnetic easy-axis. In contrast, along the x-direction, $\vv{M}_\mathrm{CSB}$ increases gradually with a much higher
saturation field ($B_\mathrm{M,x}\sim 1\,\mathrm{T}$). This is a result of the gradual canting of the anti-parallel magnetizations of the CrSBr layers towards the x-direction which implies that it is an in-plane magnetic hard-axis. The crystal also has an out-of-plane magnetic anisotropy, with the z-axis as the hardest magnetic axis (see SI, sec. 14). When graphene is brought on top of the CrSBr flake, the magnetic behavior of the outermost CrSBr layer gets imprinted in the graphene so that the magnetization of graphene ($\vv{M}_\mathrm{Gr}$) is expected to be collinear to the magnetization of the outermost layer of the CrSBr flake (the alignment of $\vv{M}_\mathrm{Gr}$ and $\vv{M}_\mathrm{CSB}$ is further discussed in SI, sec. 13). In the vdW heterostructure the dangling bond-free surface of CrSBr and its atomic flatness reduce corrugations and roughness, leading to a boost in the proximity effects. 

\begin{figure*} [!htb]
\centering\includegraphics[width=0.85\textwidth]{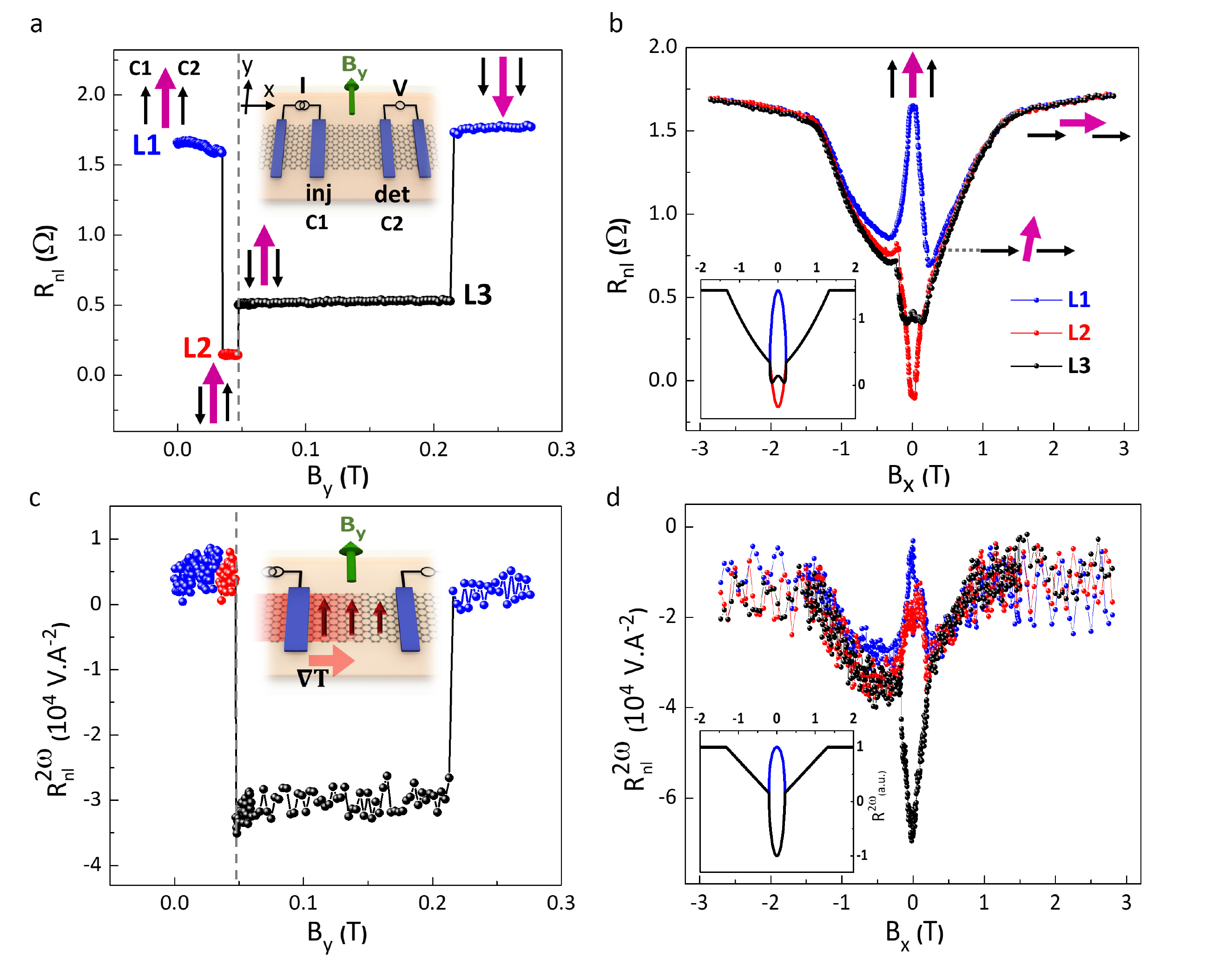}
\caption{\textbf{Electrical and thermal generation of spin currents in the magnetized Gr.} (a) Non-local spin-valve measurement; 1\ts{st} harmonic non-local resistance $R_\mathrm{nl}=V/I$ versus $B_\mathrm{y}$ (with $I=5\,\mu$A), showing three levels (L1, L2 and L3) corresponding to the different configurations of Co injector$/$detector magnetization directions ($\vv{M}_\mathrm{Co}$: black arrows) with respect to that of CrSBr ($\vv{M}_\mathrm{CSB}$: purple arrows). Inset: Schematic of the non-local measurement geometry, with C1 and C2 Co electrodes as the spin injector and detector, respectively. (b) Modulation of $R_\mathrm{nl}$ as a function of $B_\mathrm{x}$, measured in Hanle geometry with initial alignment of $\vv{M}_\mathrm{Co}$ of injector and detector and $\vv{M}_\mathrm{CSB}$ (corresponding to L1, L2 and L3, defined in panel a). The asymmetry in the Hanle curves could be related to the possible few-degrees misalignment of the magnetization axes of the Co electrodes with respect to that of CrSBr crystal and/or misalignment of $B$ from the x-axis.  Inset: Hanle curves calculated by the spin-dependent graphene conductivity model. The best fit to the measured results is provided considering $P_\mathrm{Gr}\approx 14 \,\%$ and $P_\mathrm{inj} \approx P_\mathrm{det} \approx -24\,\%$ (see SI, sec. 9). (c,d) Spin-dependent Seebeck effect; (c) 2\ts{nd} harmonic spin-valve measurement of $R_\mathrm{nl}^{2\omega}=V^{2\omega}/I^2$ versus $B_\mathrm{y}$. Inset: Schematic of the non-local measurement geometry, showing temperature gradient ($\nabla T$) in the graphene channel due to Joule heating at the current source contacts that results in the thermal generation of spin current due to the finite $P_\mathrm{Gr}$. (d) Modulation of the $R_\mathrm{nl}^{2\omega}$ as a function of $B_\mathrm{x}$ in the Hanle geometry. Inset: Theoretically calculated Hanle curves for the 2\ts{nd} harmonic signal, considering the SdSE as the spin generating mechanism (see SI, sec. 12). All measurements are performed at $T=4.5\,$K.}
    \label{fig:two}
\end{figure*}

The presence of spin-dependent conductivity in graphene is directly observed 
 in the spin-valve design shown in Figure 1c. Using the three-terminal (3T) geometry, the resistance is measured versus the magnetic field $B_\mathrm{y}$, applied along the easy axis of the Co electrodes which is the same as that of the CrSBr crystal. As shown in panel d, in the spin-valve heterostructure of $\mathrm{Co/Al_2O_3/magnetic\,Gr}$, the relative orientation of $\vv{M}_\mathrm{Gr}$ with respect to $\vv{M}_\mathrm{Co}$, defines the spin polarization of the injected current and therefore considerably changes the resistance of the contact between graphene and Co ($R_\mathrm{3T}$). 
The abrupt change in resistance, depending on the relative orientation of $\vv{M}_\mathrm{Gr}$ and $\vv{M}_\mathrm{Co}$ resembles the GMR effect, but with the advantage of the long-distance spin transfer by the graphene. This would not happen in a conventional 3T measurement with a non-magnetic graphene, because as the spin injection changes sign when the Co magnetization is reversed, simultaneously the sign of the detection is also reversed, resulting in no magnetoresistance \cite{dash2009electrical}. Therefore 
the observation of the 3T spin-valve effect is only possible if the graphene is magnetic, with spin polarization of conductance $P_\mathrm{Gr}$. The change in the $R_\mathrm{3T}$ versus $B_\mathrm{y}$ is proportional to $P_\mathrm{Gr}$ ($\Delta R_\mathrm{3T}\propto P_\mathrm{Gr}$, see SI sec. 6).

\begin{figure*} [!htb]
\centering\includegraphics[width=0.93\textwidth]{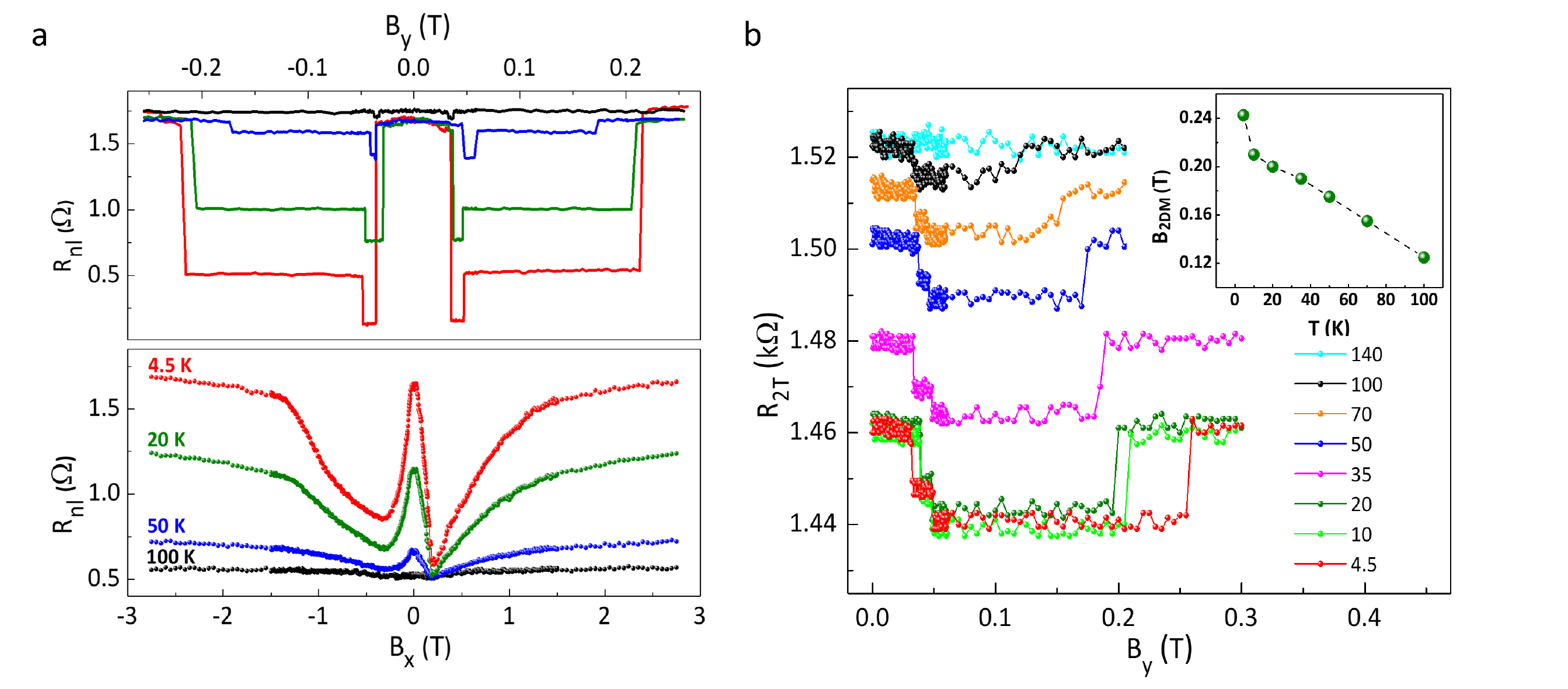}
\caption{\textbf{Temperature dependence of the spin signal.} (a) The non-local spin-valve and Hanle measurements (with the parallel initial configuration of $\vv{M}_\mathrm{CSB}$, $\vv{M}_\mathrm{Co,inj}$ and $\vv{M}_\mathrm{Co,det}$) at various temperatures. The measured spin-valve data is shifted along the y-axis for a more clear demonstration (see SI sec.~3, for the $R_{\mathrm{nl}}$ shown without the offset). (b) Two-terminal resistance ($R_\mathrm{2T}$) measured between contacts C1 and C2 versus $B_{\mathrm{y}}$, at various temperatures.  Inset: The gradual decay in the switching field of CrSBr along its easy axis ($B_\mathrm{M,y}$) by the increase in temperature. This is consistent with the temperature dependence of the SQUID magnetometry of CrSBr \cite{telford2020layered}.}
    \label{fig:three}
\end{figure*}

We study this spin-charge current coupling in graphene further by measuring the pure spin current generated by the magnetized graphene in the non-local four-terminal (4T) geometry of Figure 2a, where the charge current path can be fully separated from the voltage detection circuit. The non-local resistance ($R_\mathrm{nl}=V_\mathrm{nl}/I$) is measured  versus $B_\mathrm{y}$. In an equivalent measurement on pristine graphene one would observe \emph{two} (non-local) resistance levels associated to the parallel and anti-parallel magnetization alignment of the injector and detector Co electrodes \cite{tombros2007electronic}. In contrast, here we observe \emph{three} resistance levels that are only possible if the spin transport in graphene depends on the relative orientation of $\vv{M}_\mathrm{Gr}$ (or $\vv{M}_\mathrm{CSB}$) with respect to the magnetization of the injector $\vv{M}_\mathrm{Co,inj}$ and that of the detector $\vv{M}_\mathrm{Co,det}$ electrodes. The spin-valve measurement is performed with initial alignment of all three magnetic elements (injector, detector and CrSBr) at $B_\mathrm{y}= -1\,$T. By increasing the field $B_\mathrm{y}$ starting from 0 T, the magnetizations of the injector and the detector electrodes switch to the opposite direction one after the other at $B_\mathrm{y}<50\,\mathrm{mT}$. The third switch (at $B_\mathrm{y}= 0.21\,\mathrm{T}$) happens at a value of the field that is too large to be related to the Co electrodes (considering their geometrical anisotropy). However, it corresponds well with switching field of the $\vv{M}_\mathrm{CSB}$ along its easy axis, shown in the SQUID magnetometry of Figure 1b. This observation is a direct evidence for the non-zero spin polarization of the graphene conductivity as a result of the proximity-induced magnetism.

The spin polarization in the magnetized graphene is defined as $P_\mathrm{Gr}=(\sigma_\mathrm{u}-\sigma_\mathrm{d})/(\sigma_\mathrm{u}+\sigma_\mathrm{d})$ with $\sigma_\mathrm{u}$ and $\sigma_\mathrm{d}$ as conductivity for spin-up and spin-down channels, respectively. Considering the spin-diffusion in the graphene channel and solving the coupled spin-charge transport equations, we derive (SI, sec. 6) the non-local resistance as:
\begin{align*}
{R_\mathrm{nl}= \dfrac{\lambda R_\mathrm{sq}}{2W} e^{-L/\lambda} (P_\mathrm{inj}-P_\mathrm{Gr}) (P_\mathrm{det}-P_\mathrm{Gr})}
\end{align*}
where $P_\mathrm{inj}$ and $P_\mathrm{det}$ are the spin polarizations of the injector and detector contacts and $\lambda$, $R_\mathrm{sq}$, L and W are the spin relaxation length, square-resistance, length and width of the graphene channel (in between the injector and detector), respectively. 
We expand the above expression for the non-local signal in order to interpret each term.
\begin{align*}
R_\mathrm{nl}\propto P_\mathrm{inj}*P_\mathrm{det}-P_\mathrm{Gr}*P_\mathrm{det}-P_\mathrm{inj}*P_\mathrm{Gr}+P_\mathrm{Gr}*P_\mathrm{Gr}
\end{align*}
$P_\mathrm{inj}*P_\mathrm{det}$ term corresponds to a regular spin diffusion in graphene, i.e. spin signal injected/detected via ferromagnetic injector/detector contacts. $P_\mathrm{inj}*P_\mathrm{Gr}$ is due to the regular spin injection via ferromagnetic contact, but the non-local signal is detected as charge voltage that builds up due to spin-to-charge conversion that happens in graphene itself. $P_\mathrm{Gr}*P_\mathrm{det}$ corresponds to the generation of spin current in graphene itself which is detected as a built up spin accumulation by ferromagnetic detector. Finally, $P_\mathrm{Gr}*P_\mathrm{Gr}$ corresponds to the spin signal that is both generated and detected by graphene itself. The presence of this term implies that in principle spin polarized contacts are not required in order to observe the graphene charge-spin current coupling. However since this term is constant versus the applied magnetic field it might not be possible to differentiate it from a spurious background when only non-magnetic contacts are used. Here we estimate $\lambda$ to be about $630\,$nm and obtain the polarizations as $P_\mathrm{Gr}\approx 14 \,\%$ and $P_\mathrm{inj} \approx P_\mathrm{det} \approx -24\,\%$ (SI, sec. 4 to 10). 
Having the $P_\mathrm{Gr}$, we roughly estimate the exchange splitting to be $\Delta=2 E_\mathrm{F}*P_\mathrm{Gr}\sim\,$20 meV that corresponds to $B_\mathrm{exch}\sim 170\,$T, assuming $E_\mathrm{F}$ to be the same as in device D2 (see SI, sec. 11).

\begin{figure*}  [!htb]
\centering\includegraphics[width=0.9\textwidth]{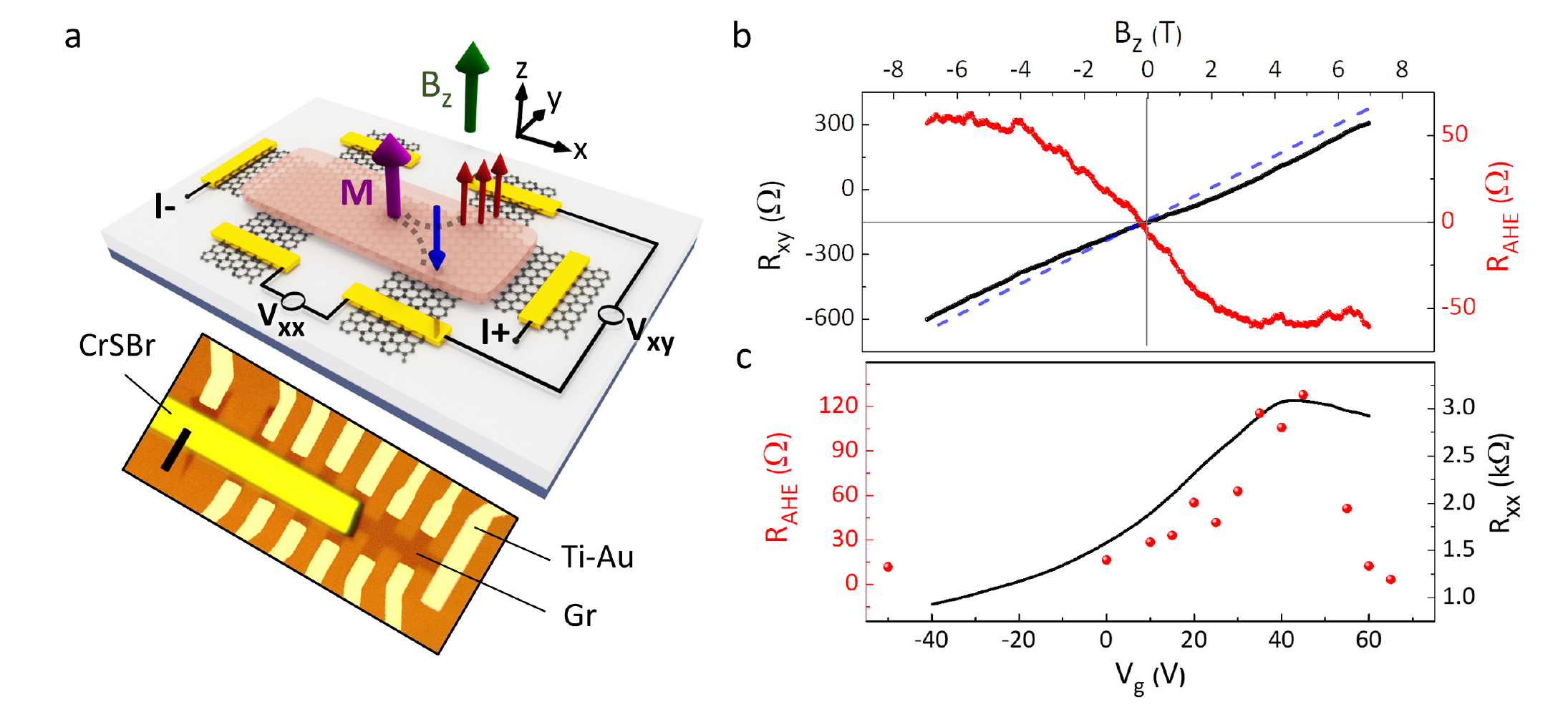}
\caption{\textbf{Anomalous Hall effect (AHE) in a Gr/CrSBr vdW heterostructure.} (a) An optical micrograph and schematic of device D2 consisting of a graphene Hall bar on $\mathrm{SiO_2}$, partially covered with a CrSBr flake (thickness $\approx$ 50 nm) with Ti(5 nm)/Au (100 nm) electrodes. Scale bar: 5 $\mu$m. The red and blue arrows represent the out-of-plane spins. (b) Left axis: Transverse resistance $R_\mathrm{xy}=V_\mathrm{xy}/I$ as a function of an out-of-plane magnetic field ($B_\mathrm{z}$), measured at $V_\mathrm{g}=+35\,$V, with $I=0.5\,\mu$A. The dashed blue line determines the linear background attributed to ordinary Hall effect. Right axis: The non-linear component of the $R_\mathrm{xy}$, attributed to the AHE ($R_\mathrm{AHE}$). (c) Left axis: The magnitude of the AHE signal (defined as an average of the maximum and minimum values of the $R_\mathrm{AHE}$ at $B_\mathrm{z}>+4\,$T and $<-4\,$T) at the various $V_\mathrm{g}$, shown by the red dots. Right axis: Longitudinal resistance ($R_\mathrm{xx}$) versus back-gate voltage ($V_\mathrm{g}$), shown by the black line. The measurements are performed at $T=30\,$K. }
    \label{fig:four}
\end{figure*}

The modulation of the spin signal under the magnetic field $B_\mathrm{x}$ applied in plane, perpendicular to the Co and CrSBr easy-axis, further confirms the presence of the very large $B_\mathrm{exch}$ in the graphene (Figure 2b). The $B_\mathrm{x}$ pulls both $\vv{M}_\mathrm{Co}$ and $\vv{M}_\mathrm{CSB}$ towards the x-axis, with saturation fields of $B_\mathrm{Co}\approx 0.2\,\mathrm{T}$ and $B_\mathrm{CSB}\approx 1.3\,\mathrm{T}$. We label this experiment a Hanle measurement since the $B_\mathrm{x}$ is applied perpendicular to the initial direction of the injected spins, even though there seems to be no spin precession involved in contrast with conventional Hanle precession measurements. The Hanle curves in Figure 2b are measured after initial alignment of the $\vv{M}_\mathrm{Co}$ and $\vv{M}_\mathrm{CSB}$ along the y-axis, setting the magnetization configuration corresponding to each resistance level of the spin-valve measurement (L1, L2 and L3 in panel a). For L1, the $R_\mathrm{nl}$ has its maximum value at $B_\mathrm{x}=0\,$T, as the injected spins are aligned with the $\vv{B}_\mathrm{exch}$. However, once the $B_\mathrm{x}$ pulls the $\vv{M}_\mathrm{Co}$ along the x-direction, the strong $\vv{B}_\mathrm{exch}$ fully randomizes the component of the injected spins that are perpendicular to it. At $B_\mathrm{x}=0.2\,$T, the magnetization directions of contacts are saturated along the x-axis while $\vv{B}_\mathrm{exch}$ is still mostly pointing along the y-axis. This yields to only a small projection of the injected spins in the direction of $\vv{B}_\mathrm{exch}$, thus resulting in a considerable decay of $R_\mathrm{nl}$.
The increase of $B_\mathrm{x}>0.2\,$T pulls the $\vv{M}_\mathrm{CSB}$ further along the x-direction with saturation at 1.3 T when the $\vv{M}_\mathrm{CSB}$ aligns with $\vv{M}_\mathrm{Co}$ once again and thus $R_\mathrm{nl}$ retrieves its initial value. The Hanle curves measured for the anti-parallel alignment of $\vv{M}_\mathrm{CSB}$ with respect to $\vv{M}_\mathrm{Co,inj}$ or $\vv{M}_\mathrm{Co,det}$ (corresponding to level L2 or L3 in panel a) also show similar behavior, but with the distinct initial value of $R_\mathrm{nl}$ at $B$= 0 T. The overall behavior of the spin signal in the Hanle curves is not determined by the spin-precession, but rather by the relative orientation of the Co and graphene magnetizations. This is due to the very large $B_\mathrm{exch}$ which allows for the information transfer only by the spins collinear to the $\vv{M}_\mathrm{Gr}$. In the inset of Figure 2b (bottom left) we show the Hanle curves derived from the analytic expression for the $R_\mathrm{nl}$ (SI, sec. 8) which agree well with the experimental results.

The strong induced magnetism in graphene also leads to the unprecedented observation of spin-dependent Seebeck effect (SdSE) \cite{uchida2008observation, rameshti2015spin}. Due to the spin dependence of the Seebeck coefficient, we can generate spin current by having a thermal gradient in the magnetized graphene channel. We measure the 2\ts{nd} harmonic signal that is associated with thermal effects due to Joule heating ($\Delta T \propto I^2$). Figure 2c shows that the non-local  2\ts{nd} harmonic resistance ($R_\mathrm{nl}^{2\omega}=V^{2\omega}/I^2$) abruptly changes with the switch in the detector magnetization direction (at $B_\mathrm{y}\sim 50\,$mT) getting anti-parallel to the graphene magnetization. The spin signal retrieves its initial value when $\vv{M}_\mathrm{CSB}$ also switches (at $\sim 0.21\,$T) and gets parallel to $\vv{M}_\mathrm{Co,det}$ again. We observe that the switch in the direction of the injector magnetization at $B_\mathrm{y}=35\,$mT does not change the $R_\mathrm{nl}^{2\omega}$. This assures the thermal origin of the measured spin signal that is generated only by the Joule heating of graphene at the injector contact, independent of the injector magnetization. This is the first observation of the thermal generation of spin accumulation by a magnetized graphene, attributed to the SdSE. In Figure 2d, we demonstrate the modulation of the SdSE spin signal versus $B_\mathrm{x}$ that is measured for the magnetization configurations of the injector, detector and CrSBr (defined for L1, L2 and L3 in panel a). The modulation of the spin signal versus $B_\mathrm{x}$ in the 2\ts{nd} harmonic is 
understood considering the collinearity of the thermally injected spins with the magnetization of graphene, consistent with theoretically calculated curves shown in the inset (also see SI, sec. 12). The similar behavior of the 2\ts{nd} harmonic L1 and L2 Hanle curves further confirms that $\vv{M}_\mathrm{Co,inj}$ has  no influence on the detected signal.

The generation of the spin currents by the magnetized graphene should persist up to the relatively high N\'eel temperature of CrSBr ($T_\mathrm{N}\approx$ 132 K). We examine this by the local and non-local spin transport measurements at various temperatures (Figure 3). The dependence of the spin signal on temperature reflects the temperature-dependence of the magnetization of CrSBr layers.
The spin-valve and Hanle curves, measured non-locally up to $T$= 100 K (panel a) show considerable decay of the spin signal. 
Such decay is attributed to the randomization of the magnetization of CrSBr, since the decay is much larger than what is expected from the temperature dependence of Co spin polarization or spin transport in graphene on non-magnetic substrates \cite{villamor2013temperature, Han2014}. This is further confirmed by two-terminal (2T) measurements of Figure 3b. The considerable spin polarization of graphene in this system 
allows for detection of the large spin signal in the local 2T geometry (measured with contact C1 and C2). As expected, each spin-valve measurement contains three switches that are corresponding to the magnetization switch of C1, C2 and CrSBr. Consistent with the non-local measurements, the size of the spin-valve switches shrink with the increase in temperature and fully vanishes (below the noise level) at $T>100\,$K. Moreover, we observe that as the temperature increases, the magnetic field at which $\vv{M}_\mathrm{CSB}$ switches along its easy axis ($B_\mathrm{M,y}$) shifts towards smaller values (shown in the inset). This also indicates suppression of the induced magnetism in graphene as the temperature reaches the critical value ($T_\mathrm{N}$), consistent with the temperature-dependence of the SQUID magnetometry of CrSBr \cite{telford2020layered}.

Another significant and direct consequence of the induced magnetism in graphene is the emergence of the anomalous Hall effect that is subject to the co-presence of SOC and magnetism \cite{nagaosa2010anomalous}. We assess this by interfacing a thin exfoliated CrSBr bulk flake with a graphene Hall bar 
(Figure 4a) and measuring the transverse voltage ($V_\mathrm{xy}$) as a function of the out-of-plane magnetic field ($B_\mathrm{z}$) when longitudinal current is applied. The $B_\mathrm{z}$ gradually pulls the magnetization of the outermost layer of CrSBr (and so the magnetization of graphene) out of the 2D plane leading to the imbalance in density of the out-of-plane spins.  The AHE introduces sizable non-linearity in the B-dependence of the transverse resistance $R_\mathrm{xy}$ shown in Figure 4b. The subtraction of the ordinary Hall effect (linear in B) provides us with the solitary contribution of AHE ($R_\mathrm{AHE}$) that saturates at $B_\mathrm{z}\sim\,4\,$T. The strength of the AHE depends on the position of the Fermi-energy in the band structure of the magnetized graphene. Figure 4c shows an increase in the AHE signal as the Fermi-energy approaches the charge neutrality point, reaching the maximum of $R_\mathrm{AHE}\approx\,$120$\,\Omega$ (measured at $T=30\,$K) and preserves sign for both electrons and holes. The magnitude of the AHE signal is comparable with that reported for graphene on YIG \cite{wang2015proximity, tang2018approaching}. However, note that in this geometry (of device D2) graphene lies on the $\mathrm{SiO_2}$ substrate that causes corrugations in graphene and so reduces its interaction with the CrSBr flake that is later transfered on top. Therefore it is expected to have larger AHE signal, if CrSBr is used as a substrate (similar to device D1). The observation of the AHE in graphene/CrSBr vdW heterostructures not only confirms the induced magnetism but also indicates enhanced SOC in the graphene. The induced SOC allows for the emergence of spin-to-charge conversion mechanisms \cite{mendes2015spin}, offering even broader range of applications.

These findings present the air-stable graphene/CrSBr vdW heterostructure as the excellent choice where magnetism, spin-orbit coupling and the long spin lifetime are brought together in a 2D carbon lattice. Graphene with the high sensitivity of charge and spin transport to the magnetization of the outermost layer of the neighboring 2D AFM CrSBr, provides a tool for studying the behavior of a single magnetic sub-lattice. The direct measurement of the spin polarized conductance of magnetic graphene, opens the platform for its implementation in the prospective 2D memory technology with ultra-fast operation and long-distance transfer of spin information.  
The unprecedented electrical and thermal generation of spin currents by the magnetized graphene allows for the design of magneto-electric devices (such as spin-valves) without the need for magnetic injector/detector electrodes which leads to 
substantial advances in the 2D spintronics and caloritronics.  These findings not only have numerous applications, but also have fundamental significance in realization of the physics of magnetism in graphene and the modulation of its spin texture.  \\

\textbf{Methods}

$Device\,Fabrication$.
The bilayer graphene and CrSBr flakes are mechanically cleaved from their bulk crystals on $\mathrm{SiO_2}$/Si substrates, using adhesive tapes \cite{novoselov2005two}. The bilayer graphene flakes are identified by their optical contrast with respect to the substrate \cite{li2013rapid}. The thicknesses of the flakes are verified by atomic force microscopy.\
Device D1; Using a dry pick-up technique \cite{Zomer2014}, we transfer the bilayer graphene on the bulk CrSBr flake by a Polycarbonate(PC)-PDMS stamp. The PC is removed in chloroform, followed by annealing in Ar/$\mathrm{H_2}$ atmosphere for 6 hours at 350 C. The preparation of the vdW stack is followed by the fabrication of $\mathrm{Al_2O_3}\,(0.9\,\mathrm{nm})/\mathrm{Co}$(30 nm) electrodes on the vdW stack by e-beam lithography technique (using PMMA as the e-beam resist). \ 

Device D2 (AHE); The large area bilayer graphene flake on the $\mathrm{SiO_2}$ substrate is initially etched into a Hall-bar geometry in $\mathrm{O_2}$ plasma environment, using a pre-patterned PMMA-membrane as the mask. The etching procedure is followed by mechanical removing of the PMMA membrane, leaving the surface of the graphene flake residue free. The bulk CrSBr flake, initially exfoliated on a PDMS stamp is transferred on the etched graphene, partially covering the Hall bar. Device fabrication is completed by the e-beam lithography of the  Ti(5 nm)/Au(100 nm) electrodes, deposited by the e-beam evaporation of the metals in ultra-high vacuum. \\

$Electrical\,Measurements$. The charge and spin transport measurements are performed by using a standard low-frequency ($<$ 20 Hz) lock-in technique with AC current source up to 10 $\mu$A. A Keithley source-meter is used as the DC-voltage source for the gate. Rotatable sample stages (separate for the in-plane and out-of-plane measurements) are used for applying the magnetic field by a (superconducting) magnet in all the possible directions. \\

$SQUID\,Measurements$.  The DC magnetic susceptibility is measured in a Cryogenic R-700X SQUID magnetometer. A single crystal of CrSBr is loaded in a determined orientation with respect to the applied magnetic field and the magnetization was measured as a function of applied magnetic field. \\

$CrSBr\,Synthesis$. CrSBr single crystals are synthesized using a modified chemical vapor transport approach adapted from the original report by Beck \cite{beck1990chalkogenidhalogenide} as described in ref \cite{telford2020layered}. Disulfur dibromide and chromium metal are added together in a 7:13 molar ratio to a fused silica tube approximately 35 cm in length, which is sealed under vacuum and placed in a three-zone tube furnace. The tube is heated in a temperature gradient (1223 to 1123 K) for 120 hours. CrSBr grows as black, shiny flat needles along with $\mathrm{CrBr_3}$ and $\mathrm{Cr_2S_3}$ as side products. CrSBr crystals are cleaned first by washing in warm pyridine, water, and acetone, and then by mechanical exfoliation to ensure no impurities remained on the surface.\\



\newpage
\textbf{Acknowledgement}

We would like to thank M. H. D. Guimarães and E. J. Telford for  discussions and T. J. Schouten, H. Adema, Hans de Vries and J. G. Holstein for technical support. This research has received funding from the Dutch Foundation for Fundamental Research on Matter (FOM) as a part of the Netherlands Organisation for Scientific Research (NWO), FLAG-ERA (15FLAG01-2), the European Union’s Horizon 2020 research and innovation programme under grant agreements No 696656 and 785219 (Graphene Flagship Core 2 and Core 3), NanoNed, the Zernike Institute for Advanced Materials, and the Spinoza Prize awarded in 2016 to B. J. van Wees by NWO. Synthesis, structural characterization and magnetic measurements are supported as part of Programmable Quantum Materials, an Energy Frontier Research Center funded by the U.S. Department of Energy (DOE), Office of Science, Basic Energy Sciences (BES), under award DE-SC0019443. AD is supported by the NSF graduate research fellowship program (DGE 16-44869).




\begin{thebibliography}{10}

\bibitem{baibich1988giant}
Baibich, M.~N., Broto, J.~M., Fert, A., Van~Dau, F.~N., Petroff, F., Etienne,
  P., Creuzet, G., Friederich, A., and Chazelas, J.
\newblock {\em Physical review letters}{ \bf 61}(21), 2472 (1988).

\bibitem{binasch1989enhanced}
Binasch, G., Gr{\"u}nberg, P., Saurenbach, F., and Zinn, W.
\newblock {\em Physical review B}{ \bf 39}(7), 4828 (1989).

\bibitem{slonczewski1996current}
Slonczewski, J.~C. et~al.
\newblock {\em Journal of Magnetism and Magnetic Materials}{ \bf 159}(1), L1
  (1996).

\bibitem{myers1999current}
Myers, E., Ralph, D., Katine, J., Louie, R., and Buhrman, R.
\newblock {\em Science}{ \bf 285}(5429), 867--870 (1999).

\bibitem{vzutic2004spintronics}
{\v{Z}}uti{\'c}, I., Fabian, J., and Sarma, S.~D.
\newblock {\em Reviews of modern physics}{ \bf 76}(2), 323 (2004).

\bibitem{geim2013van}
Geim, A.~K. and Grigorieva, I.~V.
\newblock {\em Nature}{ \bf 499}(7459), 419--425 (2013).

\bibitem{abergel2010properties}
Abergel, D., Apalkov, V., Berashevich, J., Ziegler, K., and Chakraborty, T.
\newblock {\em Advances in Physics}{ \bf 59}(4), 261--482 (2010).

\bibitem{Han2014}
Han, W., Kawakami, R.~K., Gmitra, M., and Fabian, J.
\newblock {\em Nat Nano}{ \bf 9}(10), 794--807 Oct  (2014).
\newblock Review.

\bibitem{gong2017discovery}
Gong, C., Li, L., Li, Z., Ji, H., Stern, A., Xia, Y., Cao, T., Bao, W., Wang,
  C., Wang, Y., et~al.
\newblock {\em Nature}{ \bf 546}(7657), 265--269 (2017).

\bibitem{gong2019two}
Gong, C. and Zhang, X.
\newblock {\em Science}{ \bf 363}(6428), eaav4450 (2019).

\bibitem{Gmitra2015}
Gmitra, M. and Fabian, J.
\newblock {\em Phys. Rev. B}{ \bf 92}, 155403 Oct  (2015).

\bibitem{garcia2018spin}
Garcia, J.~H., Vila, M., Cummings, A.~W., and Roche, S.
\newblock {\em Chemical Society Reviews}{ \bf 47}(9), 3359--3379 (2018).

\bibitem{wei2016strong}
Wei, P., Lee, S., Lemaitre, F., Pinel, L., Cutaia, D., Cha, W., Katmis, F.,
  Zhu, Y., Heiman, D., Hone, J., et~al.
\newblock {\em Nature materials}{ \bf 15}(7), 711--716 (2016).

\bibitem{wu2017magnetic}
Wu, Y.-F., Song, H.-D., Zhang, L., Yang, X., Ren, Z., Liu, D., Wu, H.-C., Wu,
  J., Li, J.-G., Jia, Z., et~al.
\newblock {\em Physical Review B}{ \bf 95}(19), 195426 (2017).

\bibitem{tang2020magnetic}
Tang, C., Zhang, Z., Lai, S., Tan, Q., and Gao, W.-b.
\newblock {\em Advanced Materials}{ \bf }, 1908498 (2020).

\bibitem{wang2015proximity}
Wang, Z., Tang, C., Sachs, R., Barlas, Y., and Shi, J.
\newblock {\em Physical review letters}{ \bf 114}(1), 016603 (2015).

\bibitem{tang2018approaching}
Tang, C., Cheng, B., Aldosary, M., Wang, Z., Jiang, Z., Watanabe, K.,
  Taniguchi, T., Bockrath, M., and Shi, J.
\newblock {\em APL Materials}{ \bf 6}(2), 026401 (2018).

\bibitem{mendes2015spin}
Mendes, J., Santos, O.~A., Meireles, L., Lacerda, R., Vilela-Le{\~a}o, L.,
  Machado, F., Rodr{\'\i}guez-Su{\'a}rez, R., Azevedo, A., and Rezende, S.
\newblock {\em Physical review letters}{ \bf 115}(22), 226601 (2015).

\bibitem{leutenantsmeyer2016proximity}
Leutenantsmeyer, J.~C., Kaverzin, A.~A., Wojtaszek, M., and Van~Wees, B.~J.
\newblock {\em 2D Materials}{ \bf 4}(1), 014001 (2016).

\bibitem{singh2017strong}
Singh, S., Katoch, J., Zhu, T., Meng, K.-Y., Liu, T., Brangham, J.~T., Yang,
  F., Flatt{\'e}, M.~E., and Kawakami, R.~K.
\newblock {\em Physical review letters}{ \bf 118}(18), 187201 (2017).

\bibitem{karpiak2019magnetic}
Karpiak, B., Cummings, A.~W., Zollner, K., Vila, M., Khokhriakov, D., Hoque,
  A.~M., Dankert, A., Svedlindh, P., Fabian, J., Roche, S., et~al.
\newblock {\em 2D Materials}{ \bf 7}(1), 015026 (2019).

\bibitem{haugen2008spin}
Haugen, H., Huertas-Hernando, D., and Brataas, A.
\newblock {\em Physical Review B}{ \bf 77}(11), 115406 (2008).

\bibitem{yang2013proximity}
Yang, H.-X., Hallal, A., Terrade, D., Waintal, X., Roche, S., and Chshiev, M.
\newblock {\em Physical review letters}{ \bf 110}(4), 046603 (2013).

\bibitem{zollner2016theory}
Zollner, K., Gmitra, M., Frank, T., and Fabian, J.
\newblock {\em Physical Review B}{ \bf 94}(15), 155441 (2016).

\bibitem{asshoff2017magnetoresistance}
Asshoff, P., Sambricio, J., Rooney, A., Slizovskiy, S., Mishchenko, A.,
  Rakowski, A., Hill, E., Geim, A., Haigh, S., Fal’Ko, V., et~al.
\newblock {\em 2D Materials}{ \bf 4}(3), 031004 (2017).

\bibitem{behera2019proximity}
Behera, S.~K., Bora, M., Chowdhury, S. S.~P., and Deb, P.
\newblock {\em Physical Chemistry Chemical Physics}{ \bf 21}(46), 25788--25796
  (2019).

\bibitem{tombros2007electronic}
Tombros, N., Jozsa, C., Popinciuc, M., Jonkman, H.~T., and Van~Wees, B.~J.
\newblock {\em Nature}{ \bf 448}(7153), 571--574 (2007).

\bibitem{fabian2007semiconductor}
Fabian, J., Matos-Abiague, A., Ertler, C., Stano, P., and {\v{Z}}uti{\'c}, I.
\newblock {\em Acta Physica Slovaca. Reviews and Tutorials}{ \bf 57}(4-5),
  565--907 (2007).

\bibitem{zhang2005experimental}
Zhang, Y., Tan, Y.-W., Stormer, H.~L., and Kim, P.
\newblock {\em nature}{ \bf 438}(7065), 201--204 (2005).

\bibitem{tse2011quantum}
Tse, W.-K., Qiao, Z., Yao, Y., MacDonald, A.~H., and Niu, Q.
\newblock {\em Physical Review B}{ \bf 83}(15), 155447 (2011).

\bibitem{zhou2010magnetotransport}
Zhou, B., Chen, X., Wang, H., Ding, K.-H., and Zhou, G.
\newblock {\em Journal of Physics: Condensed Matter}{ \bf 22}(44), 445302
  (2010).

\bibitem{chappert2010emergence}
Chappert, C., Fert, A., and Van~Dau, F.~N.
\newblock In {\em Nanoscience And Technology: A Collection of Reviews from
  Nature Journals},  147--157. World Scientific (2010).

\bibitem{behin2010proposal}
Behin-Aein, B., Datta, D., Salahuddin, S., and Datta, S.
\newblock {\em Nature nanotechnology}{ \bf 5}(4), 266--270 (2010).

\bibitem{michetti2010electric}
Michetti, P., Recher, P., and Iannaccone, G.
\newblock {\em Nano letters}{ \bf 10}(11), 4463--4469 (2010).

\bibitem{michetti2011spintronics}
Michetti, P. and Recher, P.
\newblock {\em Physical Review B}{ \bf 84}(12), 125438 (2011).

\bibitem{zollner2018electrically}
Zollner, K., Gmitra, M., and Fabian, J.
\newblock {\em New Journal of Physics}{ \bf 20}(7), 073007 (2018).

\bibitem{gibertini2019magnetic}
Gibertini, M., Koperski, M., Morpurgo, A., and Novoselov, K.
\newblock {\em Nature nanotechnology}{ \bf 14}(5), 408--419 (2019).

\bibitem{goser1990magnetic}
G{\"o}ser, O., Paul, W., and Kahle, H.
\newblock {\em Journal of magnetism and magnetic materials}{ \bf 92}(1),
  129--136 (1990).

\bibitem{qi2018electrically}
Qi, J., Wang, H., and Qian, X.
\newblock {\em arXiv preprint arXiv:1811.02674}{ \bf } (2018).

\bibitem{telford2020layered}
Telford, E.~J., Dismukes, A.~H., Lee, K., Cheng, M., Wieteska, A., Chen, Y.-S.,
  Xu, X., Pasupathy, A.~N., Zhu, X., Dean, C.~R., et~al.
\newblock {\em arXiv preprint arXiv:2005.06110}{ \bf } (2020).

\bibitem{jungwirth2016antiferromagnetic}
Jungwirth, T., Marti, X., Wadley, P., and Wunderlich, J.
\newblock {\em Nature nanotechnology}{ \bf 11}(3), 231 (2016).

\bibitem{jiang2018electric}
Jiang, S., Shan, J., and Mak, K.~F.
\newblock {\em Nature materials}{ \bf 17}(5), 406--410 (2018).

\bibitem{dash2009electrical}
Dash, S.~P., Sharma, S., Patel, R.~S., de~Jong, M.~P., and Jansen, R.
\newblock {\em Nature}{ \bf 462}(7272), 491--494 (2009).

\bibitem{uchida2008observation}
Uchida, K., Takahashi, S., Harii, K., Ieda, J., Koshibae, W., Ando, K.,
  Maekawa, S., and Saitoh, E.
\newblock {\em Nature}{ \bf 455}(7214), 778--781 (2008).

\bibitem{rameshti2015spin}
Rameshti, B.~Z. and Moghaddam, A.~G.
\newblock {\em Physical Review B}{ \bf 91}(15), 155407 (2015).

\bibitem{villamor2013temperature}
Villamor, E., Isasa, M., Hueso, L.~E., and Casanova, F.
\newblock {\em Physical Review B}{ \bf 88}(18), 184411 (2013).

\bibitem{nagaosa2010anomalous}
Nagaosa, N., Sinova, J., Onoda, S., MacDonald, A.~H., and Ong, N.~P.
\newblock {\em Reviews of modern physics}{ \bf 82}(2), 1539 (2010).

\bibitem{novoselov2005two}
Novoselov, K., Jiang, D., Schedin, F., Booth, T., Khotkevich, V., Morozov, S.,
  and Geim, A.
\newblock {\em Proceedings of the National Academy of Sciences of the United
  States of America}{ \bf 102}(30), 10451--10453 (2005).

\bibitem{li2013rapid}
Li, H., Wu, J., Huang, X., Lu, G., Yang, J., Lu, X., Xiong, Q., and Zhang, H.
\newblock {\em ACS nano}{ \bf 7}(11), 10344--10353 (2013).

\bibitem{Zomer2014}
Zomer, P.~J., Guimar{\~a}es, M. H.~D., Brant, J.~C., Tombros, N., and van Wees,
  B.~J.
\newblock {\em Applied Physics Letters}{ \bf 105}(1), 013101 (2014).

\bibitem{beck1990chalkogenidhalogenide}
Beck, J.
\newblock {\em Zeitschrift f{\"u}r anorganische und allgemeine Chemie}{ \bf
  585}(1), 157--167 (1990).

\end{thebibliography}
\end{document}